\documentclass{kapproc} 
\usepackage{graphicx}

\kluwerbib

\begin{document}



\articletitle[A Multi-Frequency Study of 3C\,309.1]{A Multi-Frequency
Study of 3C\,309.1}


\author{E. Ros
\& 
A.P. Lobanov
}
\affil{Max-Planck-Institut f\"ur Radioastronomie,
Auf dem H\"ugel 69, D-53121 Bonn, Germany}





\begin{figure}[h]
\vspace{50pt}
\includegraphics{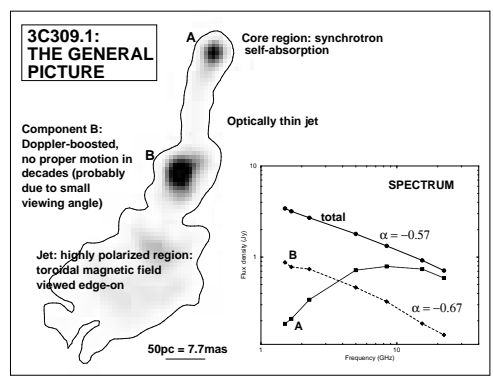}
\hfill\parbox[t]{5cm}{\caption[]{
Graphical 
summary of our VLBI results on \protect\inx{3C\,309.1} at parsec
scales in total and in linearly polarized intensity.
The linear scale is 
5.60\,$h^{-1}$\,pc\,mas$^{-1}$ for a redshift of $z$=0.905
and
$H_0$=75\,$h$\,km\,s$^{-1}$\,Mpc$^{-1}$, $q_0$=0.5.
\label{fig:ros02_01}
}
}
\end{figure}


In Figure \ref{fig:ros02_01} we summarize our results from a 
detailed multi-frequency study of
the QSO 3C\,309.1 based on the Very Long Baseline Array (VLBA)
observations made in mid 1998.  
From our images, we find a curved jet extending up to 100 milliarcseconds
(mas) to the east at low frequencies with two main components, A and B.  
A preliminary astrometric analysis (Ros and Lobanov 2001) provides a 
determination of the core position at different frequencies by 
phase-referencing to a nearby radio source, QSO S5\,1448+76.  The changes
of the core position with frequency suggest high opacity close to the 
core caused by synchrotron self-absorption.  Due to the large astrometric
uncertainties we cannot draw any conclusions about the values of the opacity
gradients at high frequencies.  We believe that a detailed analysis
of the frequency depedence of the core position will reveal the profile
of the matter distribution in the broad line region, as was initially
suggested by Lobanov (1998).


\begin{chapthebibliography}{1}
\bibitem{lob98}
A.P.\ Lobanov, A\&A, 330, 79 (1998)
\bibitem{ros01}
E.\ Ros and A.P.\ Lobanov, Opacity in the jet of 3C 309.1, in Proceedings
of the 15th Working Meeting on European VLBI for Geodesy and Astrometry, 
Rius A., Behrend D. (Eds.), IEEC, CSIC, Barcelona, 208-215 (2001)
\end{chapthebibliography}
\end{document}